\documentclass[aps,prl,twocolumn,showpacs,superscriptaddress]{revtex4}
\usepackage{graphicx}
\usepackage{epstopdf}

\begin{document}

\title{Nanoscale electronic inhomogeneity in FeSe$_{0.4}$Te$_{0.6}$ revealed through unsupervised machine learning}

\author{P. Wahl}
\email{wahl@st-andrews.ac.uk}
\affiliation{SUPA, School of Physics and Astronomy, University of St. Andrews, North Haugh, St. Andrews, Fife, KY16 9SS, United Kingdom}
\author{U. R. Singh}
\altaffiliation{Present address: Center for Hybrid Nanostructures (CHyN), Universit\"at Hamburg, Luruper Chaussee 149, 22761, Hamburg, Germany}
\affiliation{Max-Planck-Institut f\"ur Festk\"orperforschung, Heisenbergstr. 1, D-70569 Stuttgart, Germany}
\author{V. Tsurkan}
\affiliation{Center for Electronic Correlations and Magnetism, Experimental Physics V, University of Augsburg, D-86159 Augsburg, Germany}
\affiliation{Institute of Applied Physics, 5 Academiei str., Chisinau, MD 2028, Moldova}
\author{A. Loidl}
\affiliation{Center for Electronic Correlations and Magnetism, Experimental Physics V, University of Augsburg, D-86159 Augsburg, Germany}

\date{\today}

\begin{abstract}
We report on an apparent low-energy nanoscale electronic inhomogeneity in FeSe$_{0.4}$Te$_{0.6}$ due to the distribution of selenium and tellurium atoms revealed through unsupervised machine learning. Through an unsupervised clustering algorithm, characteristic spectra of selenium- and tellurium-rich regions are identified. The inhomogeneity linked to these spectra can clearly be traced in the differential conductance and is detected both at energy scales of a few electron volts as well as within a few millielectronvolts of the Fermi energy. By comparison with ARPES, this inhomogeneity can be linked to an electron-like band just above the Fermi energy. It is directly correlated with the local distribution of selenium and tellurium. There is no clear correlation with the magnitude of the superconducting gap, however the height of the coherence peaks shows significant correlation with the intensity with which this band is detected, and hence with the local chemical composition.
\end{abstract}


\maketitle

The 11 iron-chalcogenide superconductors have the simplest crystal structure of the iron-based superconductors, consisting of planar iron
layers with chalcogenide ($\mathrm{Se}$, $\mathrm{Te}$) anions above and below. The crystal structure provides a well-defined and non-polar cleavage plane between the chalcogenide layers. LEED and STM studies show no indication for a surface reconstruction\cite{tamai_strong_2010, massee_cleavage_2009}. Previous studies of the local density of states in this material by scanning tunneling microscopy have either concentrated on the superconducting state\cite{hanaguri_unconventional_2010,massee_imaging_2015,singh_evidence_2015} or not detected any electronic inhomogeneity in the energy range investigated \cite{he_nanoscale_2011}. Interest in the superconductivity in $\mathrm{FeSe}_{1-x}\mathrm{Te}_{x}$ has recently had a renaissance driven largely by the existence of topologically non-trivial surface states\cite{zhang_observation_2018} and the detection of zero bias anomalies in vortex cores\cite{zhang_observation_2018,wang_evidence_2018}. In particular for the interpretation of the latter, one of the big outstanding puzzles is why not all vortex cores exhibit zero bias anomalies, as would be expected for a topologically protected state, but only some. This hints to some influence of the chemical inhomogeneity in the material that has hitherto not been accounted for in analyzing the experiments.\\
To investigate the electronic inhomogeneity in the normal state electronic structure, we have carried out STM measurements on a single crystal of $\mathrm{FeSe}_{1-x}\mathrm{Te}_{x}$ with $x=0.61$ (determined by EDX measurement) and with a superconducting transition temperature $T_{\mathrm C}\approx 14~\mathrm{K}$\cite{tsurkan_physical_2011}. We have used a home-built low temperature STM which allows for in-situ sample transfer and cleavage\cite{white_stiff_2011}. Sample cleaving was performed at temperatures around $20~\mathrm{K}$. Spectroscopic maps in which differential tunneling conductance $\mathrm dI/\mathrm dV$ is measured as a function of bias voltage $V$ and position $\mathbf{r}$ have been acquired in the temperature range from $2~\mathrm{K}$ to $16~\mathrm{K}$ through a lock-in amplifier with a modulation of $600~{\mu\mathrm V}_\mathrm{RMS}$. The differential conductance in the normal state and superconducting state are referred to $g_\mathrm{N}(V)$ and $g_\mathrm{S}(V)$, respectively. Bias voltages are applied to the sample, with the tip at virtual ground. Tunneling spectra are acquired with open feedback loop.

Here, we employ an unsupervised machine learning approach through a cluster analysis of the tunneling spectra measured on $\mathrm{FeSe}_{0.4}\mathrm{Te}_{0.6}$. The algorithm is a variant of a $k$-means clustering algorithm (or Lloyd's algorithm). It uses a similarity analysis of spectra to categorize them, aiming to minimize the metrics defined through $\Delta(g(\mathbf x,V),g(\mathbf y,V))=\sum_i|g(\mathbf x,V_i)-g(\mathbf y,V_i)|^2$ of spectra $g(\mathbf{x},V)$ defined on a discrete lattice with voltages $V_i$. The algorithm compares individual spectra in each identified cluster to the average spectra of the cluster, and assigns them to the cluster with minimal difference. This process is performed iteratively until the clusters remain static in successive iterations. Apart from the differential conductance data $g(\mathbf{x},V)$, the only input parameter is the threshold $\overline{\Delta}$ above which spectra are considered different by the algorithm and a new cluster is created. The main difference to the $k$-means algorithm is that here, the number of clusters is not predetermined, but depends on the threshold $\overline\Delta$. Higher values of $\overline{\Delta}$ thus lead to a larger number of clusters and vice-versa.\\


\begin{figure}[!h]
\includegraphics [width=\columnwidth]{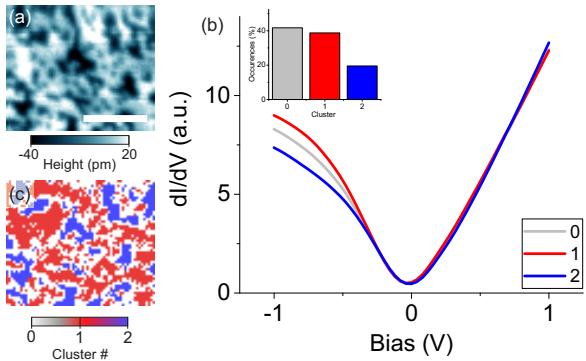}
\caption {(a) topographic STM image (scale bar: 5nm) and (b) cluster-averaged spectra, identified through the cluster algorithm described in the text for $\overline{\Delta}=7.5$, representative of Se- and Te-rich regions. Apart from a difference in differential conductance at $-1\mathrm{V}$, the spectra exhibit a small shift in the minimum close to the Fermi energy. (c) Spatial map of the cluster number spectra have been assigned to.}
\label{electronicinhomogeneity}
\end{figure}

We have applied the machine learning algorithm to two data sets to investigate spatial inhomogeneities in the normal state differential conductance $g_\mathrm N(V)$ to extract information about the normal state electronic structure of $\mathrm{FeSe}_{0.4}\mathrm{Te}_{0.6}$. The first covers a bias voltage range of $+/-1\mathrm{V}$ and the second energies in the vicinity of the Fermi energy, between $-20\mathrm{mV}$ and $10\mathrm{mV}$, i.e. in the energy range relevant for superconductivity. Application of the machine learning algorithm to the high-energy scale map reveals a dichotomy of spectra. The topography is shown in fig.~\ref{electronicinhomogeneity}(a), and the cluster-averaged spectra are plotted in panel (b). Analysis of the apparent height of topographic images yields a concentration of Se atoms of $(37\pm 4)\%$ and of Te atoms of $(63\pm 4)\%$ in the surface layer, consistent with the EDX analysis. The tunneling spectra reveal a substantial difference in differential conductance around $-1\mathrm V$ (if the tip is stabilized at $+1\mathrm V$), while above $-0.4\mathrm{V}$ only negligible differences in the shape of the spectra are found. The spatial distribution of the two most abundant spectra is in fig.~\ref{electronicinhomogeneity}(c). The spatial map reveals a stunning similarity with the topographic image shown in fig.~\ref{electronicinhomogeneity}(a), demonstrating that the two spectra identified by the machine-learning algorithm are representative of Selenium- and Tellurium-rich areas of the sample surface. A possible reason for the difference at bias voltages lower than $-0.4\mathrm{V}$ may be due to the energy of $d_z^2$-derived bands, which occur at different energies in FeTe compared to FeSe\cite{tamai_strong_2010}.\\

\begin{figure}[!h]
\includegraphics [width=8.5cm]{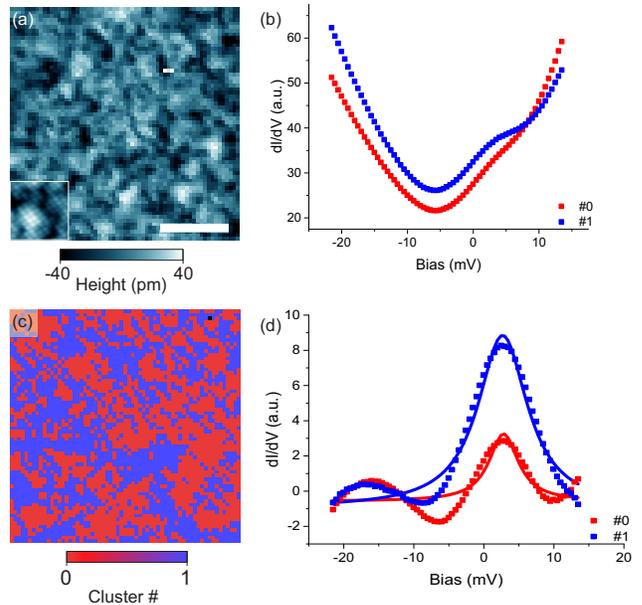}
\caption {Machine learning algorithm applied to low energy spectra. (a) Typical topography of $\mathrm{FeSe}_{0.4}\mathrm{Te}_{0.6}$ acquired simultaneously with a spectroscopic map taken in the normal state at $T=16\mathrm K$ (scale bar: 5nm), inset shows the covered area from a higher resolution topography on the same lateral and height scale, (b) Cluster-averaged spectra for Te- and Se-rich regions, respectively (obtained using $\overline{\Delta}=13$, identifying in total three clusters of spectra from the map shown in (a)). (c) Spatial map of the two most abundant clusters (the third cluster has only a single occurence, black pixel). Comparison with (a) shows that red regions tend to be Te-rich, whereas blue regions are Se-rich. (d) Same spectra as in (b), but after subtraction of a polynomial of second degree, so that the peak at $2.4\mathrm{mV}$ is better visible.}
\label{lowenergyinhomogeneity}
\end{figure}

Having demonstrated that the algorithm can extract meaningful information from spectroscopic maps, we have applied the same algorithm to investigate the low energy density of states in the vicinity of the Fermi energy in the normal state of $\mathrm{FeSe}_{0.4}\mathrm{Te}_{0.6}$, to understand the relation between the local chemical composition and the electronic states in an energy range that is relevant for superconductivity. Fig.~\ref{lowenergyinhomogeneity}(a) shows the topographic image of a differential conductance map acquired in the normal state of $\mathrm{FeSe}_{0.4}\mathrm{Te}_{0.6}$ at a temperature $T=16\mathrm K$, i.e. above the superconducting transition temperature of $T_\mathrm c\sim 14\mathrm K$. The most abundant clusters of spectra are shown in Fig.~\ref{lowenergyinhomogeneity}(b), revealing again notable differences. The spectra reveal two main differences: (1) there is a peak at an energy slightly above the Fermi energy, but within the range of the superconducting gap, that is characteristic of one cluster of spectra, but not the other, and (2) the spectra exhibit an asymmetry between positive and negative bias voltages that is different between the two clusters of spectra. Similar to the analysis performed on the large energy scale map, inspection of the spatial prevalence of the two clusters as shown in their distribution map in Fig.~\ref{lowenergyinhomogeneity}(c) shows a clear correlation between the low energy differential conductance and the chemical nature of the atoms in the surface layer. The strong variation of the intensity of the peak at positive energies is more clearly seen after subtraction of a parabolic background from the spectra, see Fig.~\ref{lowenergyinhomogeneity}(d). While Selenium-rich regions (blue areas in Fig.~\ref{lowenergyinhomogeneity}(b)) exhibit the peak in the differential conductance spectra, it becomes much weaker if not undetectable in Tellurium-rich regions (red areas in Fig.~\ref{lowenergyinhomogeneity}(b)).


\begin{figure}[!h]
\includegraphics [width=8.5cm]{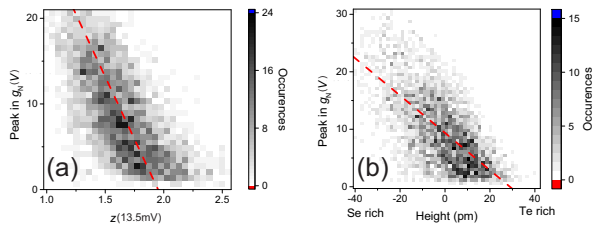}
\caption {Properties of low energy feature. (a) Correlation between the amplitude of the peak in $g_\mathrm N(V)$ and the ratio $z(V=13.5\mathrm{mV})=g(V)/g(-V)$, showing a clear anticorrelation ($C=-0.64$). (b) Correlation between the peak in the normal state differential conductance $g_\mathrm N(V)$ at $V=2.4\mathrm{mV}$ with the local apparent height as a proxy for the chemical composition. Larger heights correspond to Tellurium-rich regions and lower heights to Selenium-rich areas. An anticorrelation with a coefficient $C=-0.64$ is observed.}
\label{correlations-normalstate}
\end{figure}

We find that not only do the normal state spectra vary due to the presence or absence of the peak at $2.4\mathrm{mV}$, this is also linked to an overall asymmetry of the tunneling spectra in the range of $+15\mathrm{mV}$ and $-15\mathrm{mV}$: for tunneling spectra exhibiting a large peak at $2.4\mathrm{mV}$, the spectrum is suppressed at positive bias voltages compared to negative bias voltages, and vice-versa for spectra showing a small peak. This relation is seen in the 2D histogram shown in fig.~\ref{correlations-normalstate}(a), which shows the relation between the peak amplitude and the asymmetry of the tunneling spectra as obtained from the ratio $z(V)=g(V)/g(-V)$ for $V=13.5\mathrm{mV}$. Thus, the asymmetry in that range can serve as a proxy for the amplitude of the peak in the differential conductance.

To analyze the relation between the normal state tunneling spectra and the chemical composition more quantitatively, we show a 2D histogram between the intensity of the peak found in the normal state spectra with the local apparent height of topographic images in fig.~\ref{correlations-normalstate}(b). The apparent height is known to vary between Se and Te atoms \cite{singh_evidence_2015,aluru_atomic-scale_2019,machida_zero-energy_2019} and can hence be used as a good proxy for the local chemical composition of the top surface layer.
A clear correlation between the two is confirmed, with higher peak amplitudes found on Selenium-rich areas, and lower height in Tellurium-rich areas. The correlation coefficient is $C=-0.64$. We note that this is higher than would be expected if the local composition would change the electronic states in the iron chalcogenide layer, because that should only yield a correlation coefficient of $C=0.5$, given that the composition of only the top half the chalcogenide layer is observed, while the one of the bottom half is expected to be random. It can be argued that the correlation coefficient should be even lower if one assumes that it is the four nearest neighbour chalcogen atoms below the iron layer that need to be considered.

\begin{figure}[!h]
\includegraphics [width=8.5cm]{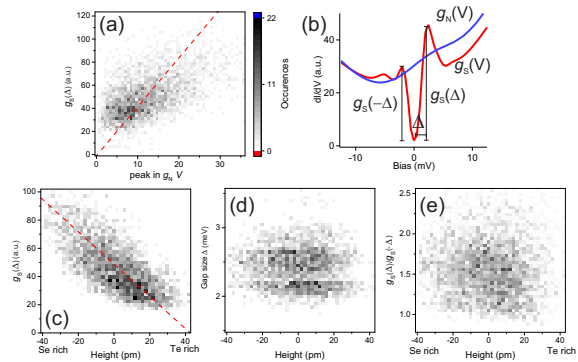}
\caption {Relation of the normal state low energy spectral feature with superconductivity. (a) Correlation between the height of the coherence peak at positive energy and the peak in the normal state differential conductance, $g_\mathrm N(V)$, showing a correlation with $C=0.54$. (b) Tunneling spectra $g_\mathrm S(V)$ and $g_\mathrm N(V)$ acquired in the normal and superconducting state, respectively, showing the definition of $\Delta$, $g_\mathrm S(\Delta)$ and $g_\mathrm S(-\Delta)$. (c) Correlation between the height of the coherence peak at positive energy and the local chemical composition, showing a correlation coefficient of $C=-0.67$. (d) 2D histogram between gap size as measured by the energy of the coherence peak at positive energy and the apparent height and (e) between the ratio of the amplitude of the coherence peaks at positive and negative energy with the apparent height, both showing no correlation ($C=-0.026$ and $C=-0.056$, respectively).}
\label{correlations-superconductivity}
\end{figure}

The question arises how this low energy electronic inhomogeneity affects superconductivity, and what the origin of the peak close to the Fermi energy is. To investigate this, we can either compare a spectroscopic map obtained in the normal state with one measured in the same location in the superconducting state. The analysis of a combination of two such maps is shown in fig.~\ref{correlations-superconductivity}(a), showing the correlation of the height of the coherence peak with the height of the peak in the normal state tunneling conductance. The histogram reveals again a clear correlation, with a correlation coefficient of about $0.5$. If we use the topographic height as proxy for the height of the peak in the normal state conductance, we find an even higher correlation of $-0.67$ (compare fig.~\ref{correlations-superconductivity}(b)). For comparison, no correlation is found between the size of the superconducting gap and the topographic height (fig.~\ref{correlations-superconductivity}(c)), consistent with previous reports\cite{singh_spatial_2013}, or the ratio in the height of the coherence peaks at positive and negative bias voltage (fig.~\ref{correlations-superconductivity}(d)).

Our analysis of the peak in the differential conductance spectra at $2.4\mathrm{mV}$ suggests that the existence of the peak is not linked to the local chemical composition. If the existence of the state was closely linked to the distribution of the Selenium and Tellurium atoms in the material, the correlation coefficient between topographic height and the peak amplitude should be significantly lower than what we observe. This is further supported by the complete insensitivity of the energy of this state to the local chemical composition deduced from topographic images. Angular resolved photoemission spectroscopy of $\mathrm{FeSe}_{0.4}\mathrm{Te}_{0.6}$ shows that there is indeed a flat band just above the Fermi energy that could be responsible for the feature we observe.\cite{okazaki_superconductivity_2014} Due to the heavy character of this band, one can expect an increase in density of states, as detected in our spectra. We note that a similar peak is observed in the normal state of LiFeAs, at an energy of $-3\mathrm{mV}$\cite{chi_imaging_2017}.\\
Thus we interpret the spatial variation of the normal state feature as effectively a result of the tunneling matrix elements between the tip and sample electronic states. This does notably not mean it is a pure surface effect, as this implies that in the bulk the interlayer coupling will be spatially modulated.
The observation of this state right above the Fermi energy, and with a similar energy as the size of the superconducting gap raises questions about how close the Cooper pairs, at least on this band, are to a BEC/BCS crossover.\cite{chubukov_superconductivity_2016} A possible indication of the proximity to the BEC/BCS crossover is the particle-hole asymmetry of the spectrum of the superconducting gap, that is quite pronouncedly shifted towards a higher coherence peak being observed at positive bias voltages. The probability of breaking a pair at positive energies will be higher, as the density of final states for this process is higher compared to negative energies, because the second electron of the pair can readily enter into the quasiparticle band. We find a ratio of the height of the coherence peaks of 1.5. We note that also other unconventional superconductors exhibit a particle-hole asymmetry in their tunneling spectra, e.g. the cuprate superconductors, though in those cases the asymmetry is due to proximity of a van Hove singularity close to the Fermi energy.\cite{fischer_scanning_2007}\\
Our work has potential implication for the interpretation of a number of recent experiments. First of all, it shows that there is a nanoscale electronic inhomogeneity, that might impact the nature of vortex core bound states in the superconducting states and might indicate one route toward the differences in the low energy electronic structure that is required to explain why only a fraction of the vortex cores in $\mathrm{FeSe}_{0.4}\mathrm{Te}_{0.6}$ exhibit zero energy states\cite{wang_evidence_2018}.\\
The variation of the normal state differential conductance has potential implications for the interpretation of measurements of the critical current in Josephson STM, as recently reported in \cite{cho_strongly_2019}. The very narrow energy interval around zero bias in which we observe variations of the tunneling matrix element in the normal state suggests that an extrapolation of the normal state resistance from outside the energy scale of the superconducting gap to estimate the normal state resistance of the junction $R_N$ is difficult and subject to spatial variations. Our data indicate a strong correlation of the height of the coherence peak as well as the peak in the normal state differential conductance with the topographic height (see figs.~\ref{correlations-normalstate}(b) and \ref{correlations-superconductivity}(b)), suggesting that the same effect may contribute to spatial variations of the critical current.\\

Our results show how unsupervised machine learning can be used to identify trends in spectroscopic STM data that would otherwise be difficult to discern. In our case, it has helped us to identify the characteristic tunneling spectra of Selenium- and Tellurium-rich regions in the iron chalcogenide superconductor $\mathrm{FeSe}_{0.4}\mathrm{Te}_{0.6}$, and has uncovered a new spectroscopic feature associated with the local chemical composition that leads to an inhomogeneity in the appearance of the superconducting gap.


\end{document}